\documentclass{iau} 
\usepackage{graphicx}

\title[Attenuation recipes for galaxies] 
{Which attenuation curves for star-forming galaxies?}

\author[V\'eronique Buat et al.]   
{V\'eronique Buat$^1$
David Corre$^1$
M\'ed\'eric Boquien$^2$
 \and Katarzyna Ma{\l}ek$^{3,1}$}

\affiliation{$^1$Aix Marseille Univ, CNRS, CNES, LAM Marseille,  France  \\[\affilskip]
$^2$Centro de Astronom\'ia (CITEVA), Universidad de Antofagasta, Avenida Angamos 601, Antofagasta, Chile \\[\affilskip]
$^3$National Centre for Nuclear Research, ul. Hoza 69, 00-681 Warszawa, Poland}

\pubyear{2015}
\volume{xxx}  
\jname{Title of your IAU Symposium}
\editors{A.C. Editor, B.D. Editor \& C.E. Editor, eds.}
\begin{document}

\maketitle

\begin{abstract}
Dust attenuation shapes the spectral energy distributions of galaxies and any 
modelling and fitting procedure of their spectral energy distributions  must account for this process. We present results of two  recent  works dedicated at measuring the dust attenuation curves in star forming galaxies at redshift from 0.5 to 3, by fitting   continuum (photometric) and line (spectroscopic) measurements simultaneously with  CIGALE  using variable attenuation laws based on  flexible recipes. 
Both studies conclude to a large variety of effective attenuation laws with an attenuation law flattening when the obscuration increases. An extra attenuation is found for nebular lines. The comparison with  radiative transfer models implies a flattening of the attenuation law up to near infrared wavelengths, which is well  reproduced  with a power-laws recipe inspired by  the Charlot and Fall recipe. Here  we  propose a global modification of the Calzetti  attenuation law to  better reproduce the results of radiative transfer models. 
\end{abstract}

\section{Introduction}
\begin{figure}
\begin{center}
 \includegraphics[width=4in]{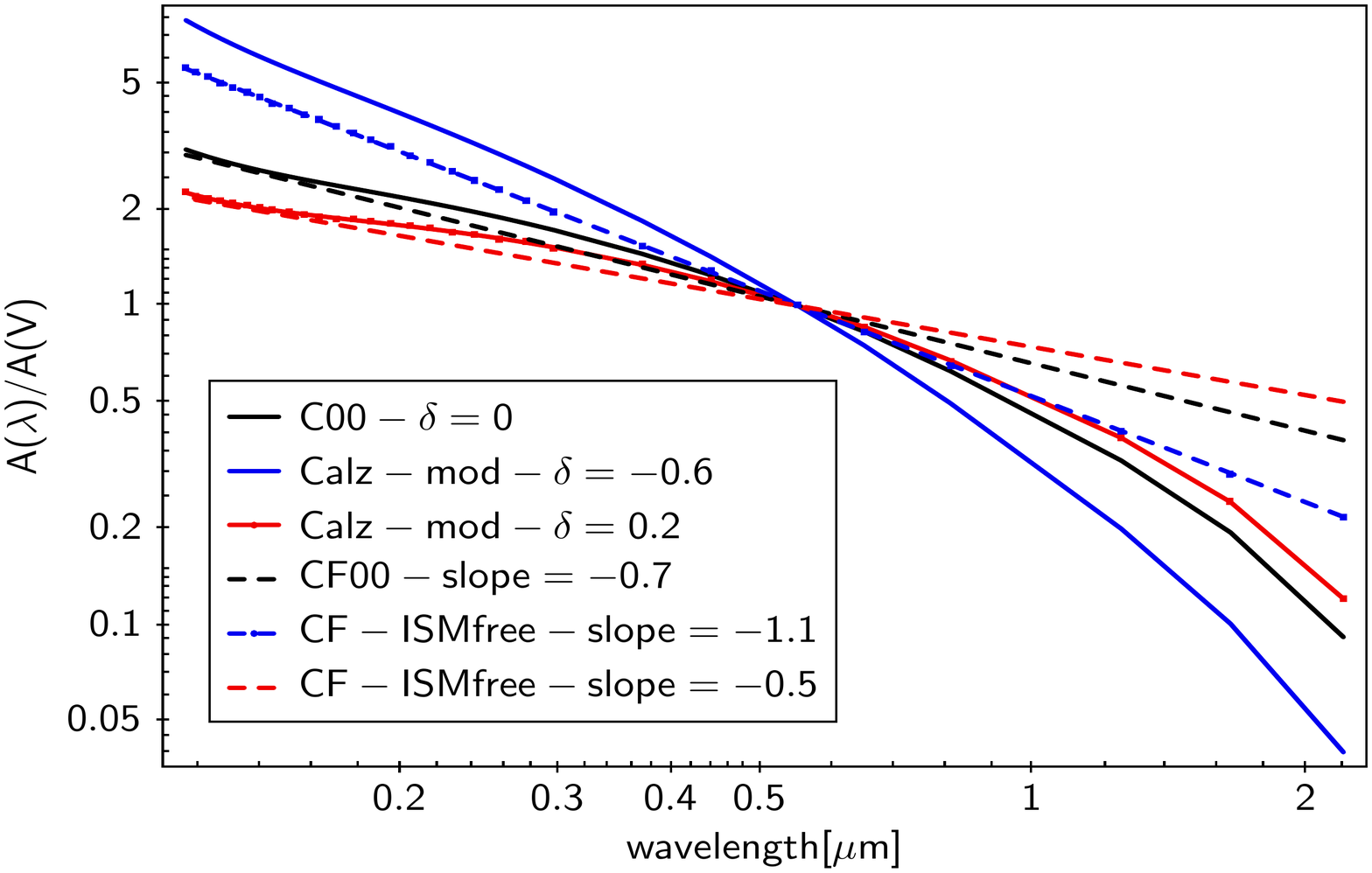} 
 \caption{Flexible effective attenuation curves based on the Calzetti et al. original law (solid lines) and on the Charlot and Fall original power-laws (dotted lines, the curves refer only here to the attenuation in the interstellar medium (ISM) and not the birth clouds). Both recipes  exhibit similar flexibility in the UV-V range. At V-NIR wavelengths the Calzetti-like recipes lead to  much steeper slopes  than the ones based on the power-laws introduced by Charlot and Fall}
   \label{fig1}
\end{center}
\end{figure}

Modelling the spectral energy distribution (SED) of galaxies is a method  commonly used to derive physical parameters useful to  quantify  galaxy evolution like stellar masses or star formation histories. Dust plays a crucial role  by strongly affecting and reshaping the spectral energy distribution (SED): it absorbs and scatters stellar photons, and thermally emits the absorbed energy in the infrared (IR) ($\lambda \sim 1-1000 ~\mu$m). 
The simplest way to model this dust effect  to introduce effective attenuation curves which account for the complex blending of dust properties and relative geometrical distribution of stars and dust within a galaxy. The most commonly used recipes are the ones proposed by \cite[Calzetti et al. (2000)]{calzetti00} (hereafter C00) and 
\cite[Charlot \& Fall (2000)]{charlot00}. 
These two recipes include a differential attenuation between young stars/nebular emission and older stellar populations. Since these original works, several authors proposed a modification of these fixed recipes (e.g. \cite[Buat et al. (2011)]{buat11}, \cite[Kriek \& Conroy (2013)]{kriek13}, \cite[Salmon et al. (2016)]{salmon16}, \cite[Lo Faro et al. (2017)]{lofaro17}). In this work we adopt flexible recipes based on both formalisms, the method is fully described in \cite[Buat et al. (2018)]{buat18}. 
The general behaviour of the two formalisms, in their flexible form, is illustrated in Fig.\,\ref{fig1}. While both recipes give similar trends in the UV-V range, the shapes of the attenuation curves are  systematically different in the V-NIR range with a flatter shape for the recipe derived from the \cite[Charlot \& Fall (2000)]{charlot00} model.

\section{Determinations of dust attenuations laws in high redshift galaxies}

We based our analysis on   two different galaxy samples.  The first one is presented in  \cite[Buat et al. (2018)]{buat18}. It  consists of galaxies in the COSMOS field,  detected with  the  PACS and SPIRE instruments of the $Herschel$ satellite and for which low resolution spectra  are  available from the 3D-HST survey. The  sample is made of  33 galaxies observed in 21 photometric bands from the NUV to the submm and with a H$\alpha$ line measurement. 
The second galaxy sample was built by \cite[Corre (2018)]{corre18} and  consists of 19  galaxies hosting a $\gamma$-ray burst (GRBHs), observed in at least 5 photometric bands starting in the UV rest frame and with H$\alpha$ and H$\beta$ line measurements.

The SEDs of all the galaxies were fitted with the CIGALE code which allows to fit simultaneously photometric and emission line fluxes (\cite[Boquien et al. (2018)]{boquien18}. Both measurements were very well fitted. The 3D-HST/COSMOS sample was fitted with both the modified C00 and Charlot and Fall recipes (\cite[Buat et al. (2018)]{buat18}). The GRBHs sample was only studied with the modified Calzetti recipe (\cite[Corre et al. (2018)]{corre18a}, \cite[Corre (2018)]{Corre18}). For the purpose of the comparison between the two studies, we focus here on the results based on the modified Calzetti recipe.

The flexible recipe is defined as:
\begin{equation}
A(\lambda) = E({\rm B-V})_{\rm star} k^{\prime}(\lambda) \left(\frac{\lambda}{\lambda_{\rm V}}\right)^{\delta},
\label{dust1}
\end{equation}
 $\delta= 0$ corresponds to the original C00 recipe. The nebular component   is extinguished with a simple screen model, a Milky Way extinction curve and a color excess $E({\rm B-V})_{\rm line}$.  The ratio $E({\rm B-V})_{\rm star}/E({\rm B-V})_{\rm line}$ is defined as a free parameter  in our recipe (it is fixed to 0.44 in the original C00 recipe).

\begin{figure}
\begin{center}
 \includegraphics[width=5in]{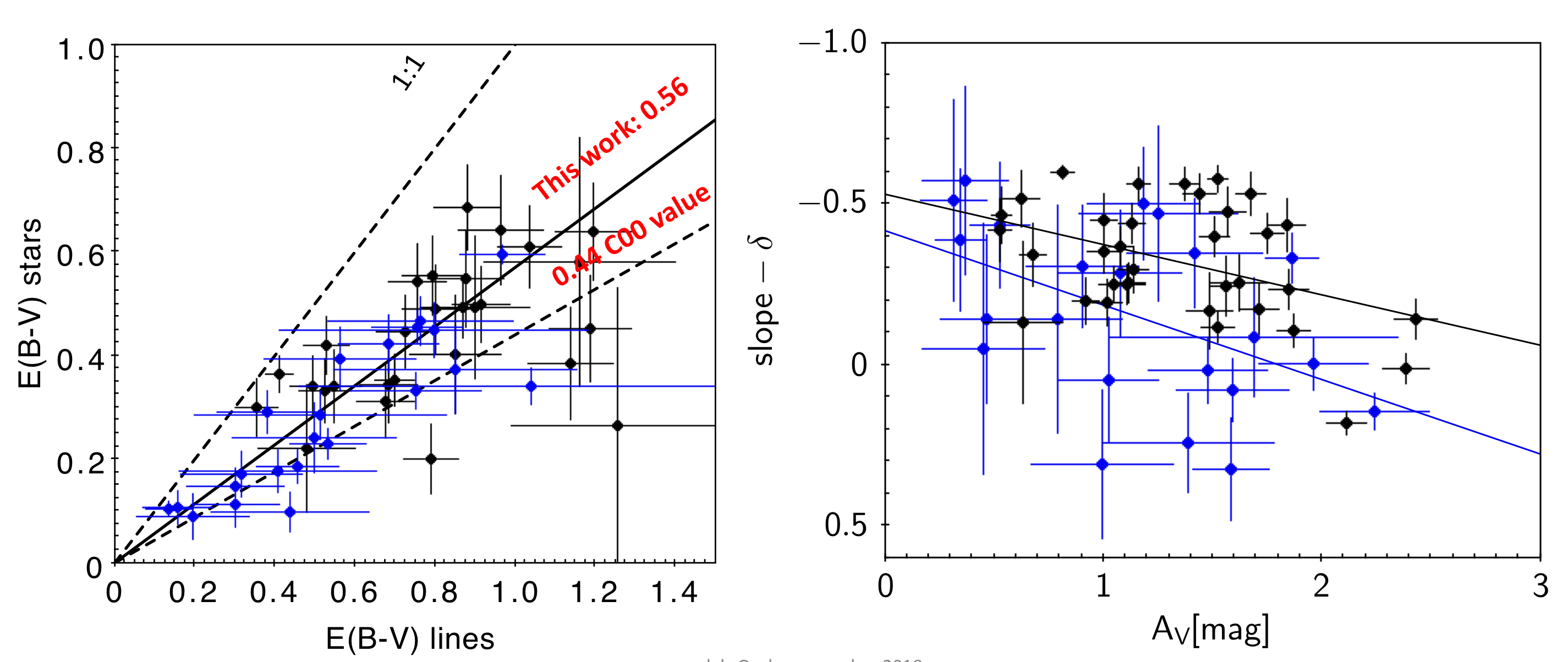} 
 \caption{Attenuation characteristics obtained from the SED fitting analysis for  the GRBH sample (blue symbols) and the 3D-HST/COSMOS sample (black symbols). Left: the color excess of the nebular emission is compared to the color excess of the stellar continuum, the average value of $E({\rm B-V})_{\rm star}/E({\rm B-V})_{\rm line}$ found for our samples and by C00 are also  plotted   as well as  as the 1:1 relation. Right: the correction $\delta$ to apply to  the original C00 attenuation curve is plotted against the global attenuation, represented by $A_{\rm V}$  }
   \label{fig2}
\end{center}
\end{figure}

The results are shown in Fig.\ref{fig2}. On the left plot, the color excesses for lines and stars  are correlated and  the two samples lead to consistent distributions. The average value of $E({\rm B-V})_{\rm star}/E({\rm B-V})_{\rm line}$ is found to be $0.56\pm 0.20$, only slightly larger than the 0.44 value  used by C00.  
The right panel of Figure 2 presents the correction $\delta$ to the slope of the attenuation law as a function of the attenuation, Av. The $\delta$ parameter spans a wide range of values, with most of the curves having steeper slopes than the original C00 law, corresponding to $\delta=0$. The majority of the 3D-HST/COSMOS galaxies is well modelled by curves with $\delta<0$, i.e. with higher levels of extinction at the blue wavelengths with respect to C00. Figure 2 also shows that, as  Av increases the attenuation law flattens, consistently with previous observational and modelling studies (e.g.  \cite[Salmon et al. (2016)]{salmon16}, \cite[Chevallard et al. (2013)]{chevallard13}). We stress that in the present study,   the shape of the attenuation curve and the relative attenuation of young and old stars are taken free parameters which was not the case in previous published works.

\section{ Consistency of  attenuation recipes with radiation transfer modelling}
 \begin{figure}
\begin{center}
 \includegraphics[width=3in]{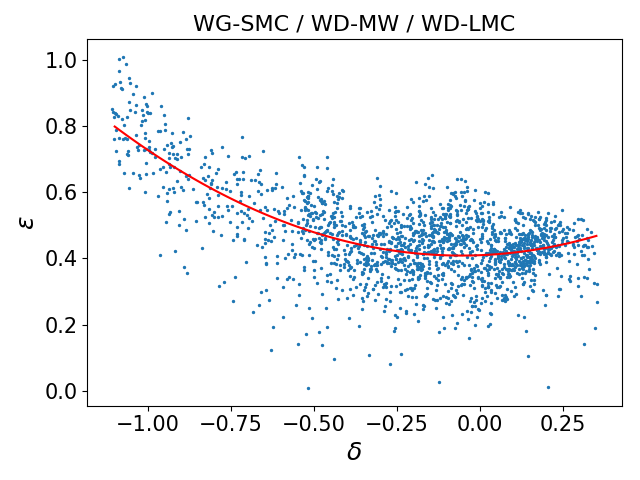} 
 \caption{Correction of the slope of the attenuation law at wavelengths larger than the V band calculated with the models of  \cite[Seon \& Draine (2016)]{seon16}. Several dust models are used, the blue points correspond to  the results of the fits performed on the numerical results  of  \cite[Seon \& Draine (2016)]{seon16}. The red curve is the best fit ($\epsilon$ as a function of $\delta$).}
   \label{fig3}
\end{center}
\end{figure}
 Radiation transfer models predict a flattening of the attenuation law when the global attenuation increase, and this flattening is seen up to the NIR. As shown in Fig.\ref{fig1}, the Calzetti recipe, modified with the introduction of $\delta$, does not allow for this flattening which is easier to get with power-laws modelling.  We propose to modify the exponent $\delta$ in $\delta+\epsilon$ at wavelength longer than the V band as follows
 
 \begin{equation}
\lambda < \lambda_{\rm V}, A(\lambda) = E({\rm B-V})_{\rm star} k^{\prime}(\lambda) \left(\frac{\lambda}{\lambda_{\rm V}}\right)^{\delta} \\
\label{dust2}
\end{equation}
 \begin{equation}
\lambda > \lambda_{\rm V}, A(\lambda) = E({\rm B-V})_{\rm star} k^{\prime}(\lambda) \left(\frac{\lambda}{\lambda_{\rm V}}\right)^{\delta+\epsilon} \\
\label{dust3}
\end{equation}
 
The value of the correction $\epsilon$ can be inferred from the radiation transfer calculations of  \cite[Seon \& Draine (2016)]{seon16}. \cite[Corre (2018)]{Corre18} fitted the numerical results of their calculations with the Calzetti modified formalism, including the $\epsilon$ correction.

 In Fig.\ref{fig3} the  values obtained of $\epsilon$ and $\delta$ are reported. The red curve is the result of  a polynomial fit:
 
  \begin{equation}
 \epsilon =  0.35~ {\delta^2}+0.04~ \delta+0.41.\\
\label{dust3}
\end{equation}

The correction  can be implemented either as this relation between $\delta$ and $\epsilon$ or as a single average value (Corre et al. 2019, in prep.).

\end{document}